\documentclass[conference, letterpaper]{IEEEtran}

\IEEEoverridecommandlockouts

\usepackage{graphicx,times,cite,amsmath, amssymb, epsfig, array, subfigure}
\usepackage{subfigure}
\usepackage[marginal]{footmisc}
\usepackage{caption}
\usepackage{algorithm}
\usepackage{balance}
\usepackage{algorithmic}
\usepackage[mathscr]{eucal}
\usepackage{cite}
\usepackage{geometry}
\renewcommand{\algorithmicrequire}{\textbf{\quad Input:}}
\renewcommand{\algorithmicensure}{\textbf{\quad Output:}}
\usepackage{multirow}

\usepackage{array}
\newlength{\figwidth}
\newcommand{\fref}[1]{Fig.~\ref{#1}}

\allowdisplaybreaks[4]
\begin{document}
\title{\LARGE{Lossy Cooperative UAV Relaying Networks: Outage Probability Analysis and Location Optimization}}
\author{
\IEEEauthorblockN{Ya Lian\IEEEauthorrefmark{1},
Wensheng Lin\IEEEauthorrefmark{1}, 
Lixin Li\IEEEauthorrefmark{1},
Fucheng Yang\IEEEauthorrefmark{2}, 
Zhu Han\IEEEauthorrefmark{3}, 
and Tad Matsumoto\IEEEauthorrefmark{4} }

\IEEEauthorblockA{\IEEEauthorrefmark{1}School of Electronics and Information, Northwestern Polytechnical University,
Xi'an, Shaanxi 710129, China}

\IEEEauthorblockA{\IEEEauthorrefmark{2}Research institute of Information Fusion, Naval Aviation University, Yantai, China}

\IEEEauthorblockA{\IEEEauthorrefmark{3}University of Houston, Houston, USA}

\IEEEauthorblockA{\IEEEauthorrefmark{4}IMT-Atlantic, France, and University of Oulu, Finland, Emeritus.}

Email: lianya@mail.nwpu.edu.cn,
\{linwest, lilixin\}@nwpu.edu.cn, 
\\fucheng85@sina.com,
zhan2@uh.edu,
tadeshi.matsumoto@oulu.fi

\thanks{This paper has been accepted for publication in IEEE Globecom 2024 workshop.}

}

\markboth{}{}
\maketitle
\begin{abstract}
In this paper, performance of a lossy cooperative unmanned aerial vehicle (UAV) relay communication system is analyzed. In this system, the UAV relay adopts lossy forward (LF) strategy and the receiver has certain distortion requirements for the received information. For the system described above, we first derive the achievable rate distortion region of the system. Then, on the basis of the region analysis, the system outage probability when the channel suffers Nakagami-$m$ fading is analyzed. Finally, we design an optimal relay position identification algorithm based on the Soft Actor-Critic (SAC) algorithm,    which determines the optimal UAV position to minimize the outage probability. The simulation results show that the proposed algorithm can optimize the UAV position and reduce the system outage probability effectively.
\end{abstract}

\section{Introduction}
In recent years, rapid development of communication technologies such as unmanned aerial vehicle (UAV) has been widespread in academia and industry\cite{Chen2024, Tu2019, DilipKumar2021}. Different from  traditional ground and fixed communication base stations (BS), UAV communication introduces mobility, which brings new challenges to the assurance of coverage and connectivity.

Cooperative relay \cite{He2019}, as a communication strategy, aims to improve the reliability of data transmission and expand the communication coverage through the cooperative operation among the network nodes including relays. 
UAV as mobile relay has attracted significant research interest. 
Chen \emph{et al.} \cite{Chen2024} characterize the outage probability of the reconfigurable intelligent surface (RIS)-equipped-UAV system under a novel modified-Fisher-Snedecor $\mathcal{F}$ fading channel model.
Tu \emph{et al.} \cite{Tu2019} derive a closed-form expression of outage probability for the wireless transmission from a BS to a mobile user via a  UAV relay over Rician fading channels. 
Dilip Kumar \emph{et al.} \cite{DilipKumar2021} derive closed-form expressions for the outage probability and throughput of a UAV-assisted full-duplex wireless system with decode-and-forward (DF) protocol over Rician fading channels. 

When UAV participates in the  communication, the performance of the communication system is likely affected by the position of UAV, so the location of UAV should be dynamically changed to satisfy different communication's Quality-of-Service (QoS) requirements, such as age-of-information (AoI) \cite{Xiao2024Statistical, Liu2023DRL-Based, Xiao2024Metaverse}. 
Presently, a lot of related research results have been published on UAV location deployment and trajectory design. 
Li \emph{et al.} \cite{Li2023trajectory} analyze the system outage probability of the multi-UAV cascaded relay communication network using orthogonal frequency division multiplexing (OFDM) and identify the optimal UAV. 
In \cite{Liu2023DRL-Based}, account is taken of the AoI as optimization objectives in UAV-enabled edge Internet-of-Things (IoT) scenarios. 
To solve the optimization problem, deep reinforcement learning (DRL) is shown to be a powerful tool.
Moreover, the impact of the UAV location on the average AoI and the peak AoI is investigated in \cite{Lin2021Cooperative} and
\cite{LIN2023249}, respectively.

The sequence may contain intra-link errors; however, the sequence is correlated to the source sequence. Hence, joint decoding at the destination may recover the original source sequence. 
This strategy is called Lossy-Forward (LF) technique\cite{Zhou2014}, which also inspires the innovation of semantic-forward \cite{Lin2024SF}. 
In terms of outage probability, LF can usually provide better performance than conventional relay technologies \cite{Lin2019lossy}. 
Due to the emergence of semantic communication \cite{Fu2024Scalable, Lin2024SIC}, lossy communication becomes promising in the future wireless networks.
Most of the literatures analyze the lossy communication performance only based on the fading model that match the fixed (on-ground) wireless communication propagation scenario \cite{Lin2019lossy, Matsumoto2024EarlyAccess}. 
Thus, it is worth conducting further research with the aim of its application to UAV aided communications. 

The contributions of this paper are summarized as follows:
\begin{itemize}
\item[$\bullet$] In this paper, by using Shannon's lossy sources-channel separation theorem, we derive the achievable rate distortion region of UAV as lossy cooperative relaying network with a specific Binary distortion requirement. 
\item[$\bullet$] Then, based on the derived region, we study the outage probability of the network under Nakagami-$m$ fading, and provide the outage derivation and the final result.
\item[$\bullet$] In addition, we design an optimal relay position identification algorithm based on the SAC algorithm to reduce the outage probability of the system while for identify the optimal relay position.
\end{itemize}

\begin{figure}[!t]
\centering \includegraphics[width=3.2in]{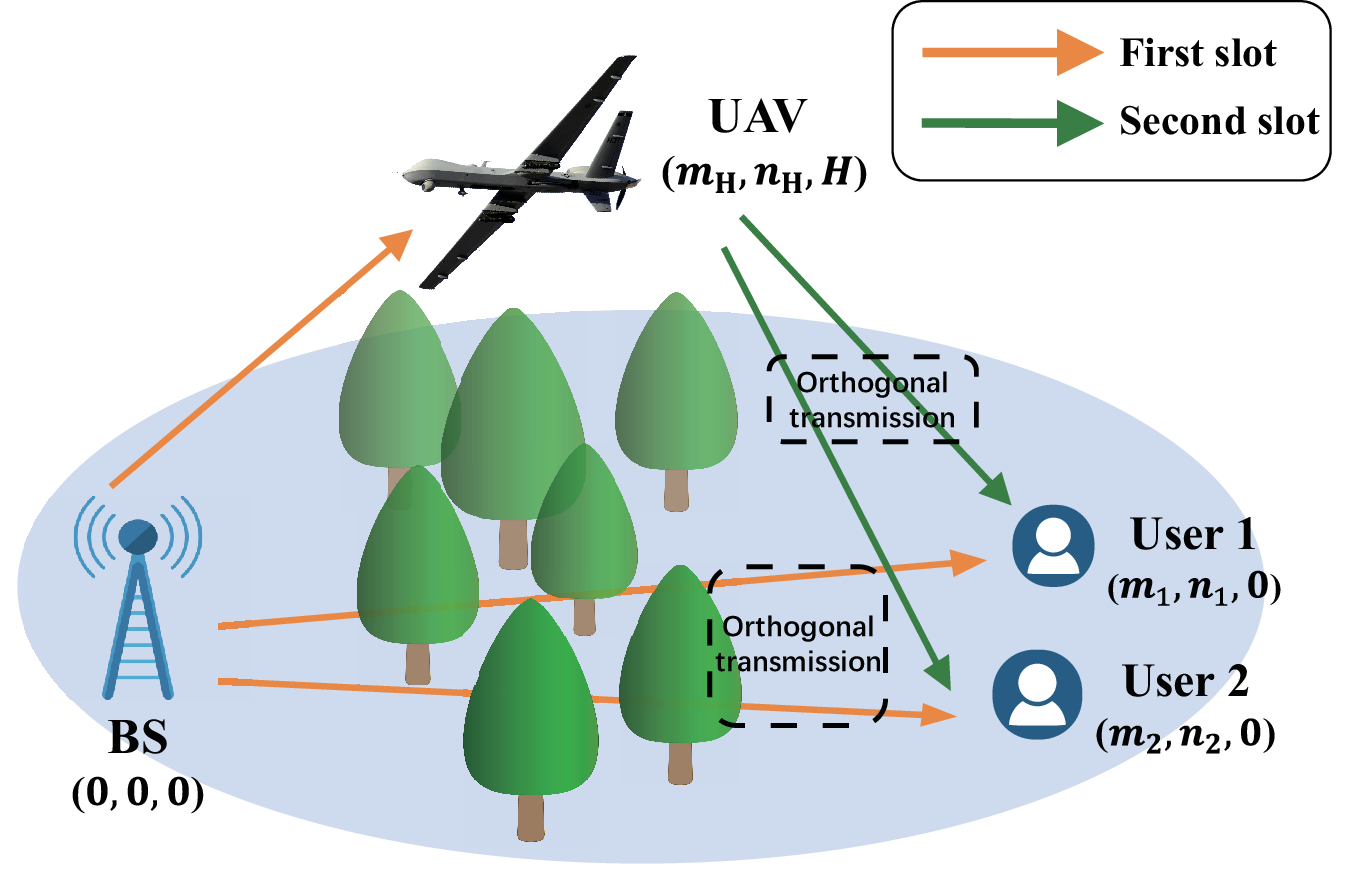}
\caption{The lossy cooperative UAV relay communication system.}
\label{fig:system}
\vspace{-1em}
\end{figure}

The rest of this paper is organized as follows. 
Section \ref{sec:system} introduces the system model of the UAV lossy cooperative ying network. 
Section \ref{sec:analysis} presents the outage probability analysis in detail. 
Section \ref{sec:sac} proposes the SAC-based algorithm to minimize the outage probability of the system. Simulation results and numerical results based on the analytical results are presented in Section \ref{sec:performance}. Finally, Section \ref{sec:conclusion} draws conclusions.

\section{Problem Statement}\label{sec:system}
In Section \ref{subsystemmodel}, we first introduce the communication model of the lossy cooperative UAV relay communication system this paper assumes in detail.
Then, Section \ref{subchannelmodel} introduces the channel model adopted in the following sections which provide theoretical results and numerical calculations.
\subsection{System Model}\label{subsystemmodel}
We consider a communication model of the lossy cooperative UAV relay communication system consisting of a BS, a UAV, and $K$ users, as shown in \fref{fig:system}. In this paper we study the case where $K=2$, i.e., two users are included, however the scenario can be extended to the case of more than two users. The users adopt orthogonal multiple access transmission (OMA) mode, e.g., frequency division multiplexing access (FDMA).
Assuming that the poor quality of the communication link between the BS and the user leads to frequent failure. To reduce the outage probability of the communication system, the UAV is located as a relay in the communication network. In lossy communication, when the BS sends the original information sequence, the user, as the receiver, has certain Binary distortion requirements as the QoS on the received information sequence. Communication interruption occurs if the distortion level of the received information does not satisfy the requirement. Since the information on sequences received by the user from the BS and from the UAV relay are correlated, the user can reduce the distortion of the received information through joint decoding after receiving these signals to achieve an acceptable degree.

In this paper, we establish the relationship between instantaneous SNR and distortion level based on Shannon's lossy source-channel separation theorem to determine the corresponding achievable link rate, given the distortion requirements.

\begin{figure}[!t]
\centering \includegraphics[width=2.7in]{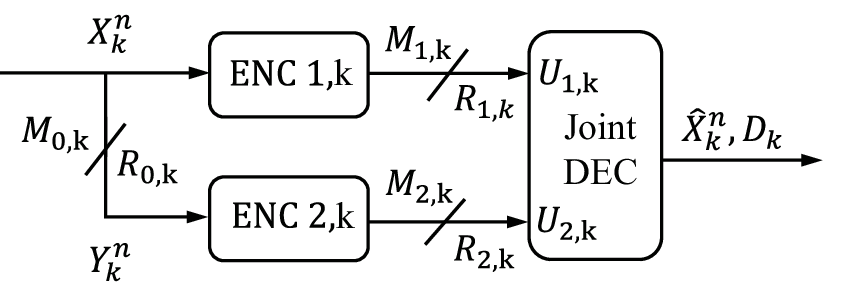}
\caption{The multiterminal source coding problem for outage probability analysis of the $k^{th}$ user.}
\label{fig:systemmodel2}
\vspace{-1em}
\end{figure}

We formulate the relationship between the distortion and each involved link's supported rate in the framework of multi-terminal source coding. As shown in \fref{fig:systemmodel2}, taking the $k^{th}$ user as an example, $X_k^n$ is equivalent to the original information sequence sent by the BS to the $k^{th}$ user, where $n$ represents the sequence length. Assuming that the information sequence is compressed into $Y_k^n$ at the bit rate $R_{0,k}$, and then the sequences $X_k^n$ and $Y_k^n$ are compressed into two code words $M_{1,k}$ and $M_{2,k}$ at the bit rates $R_{1,k}$ and $R_{2,k}$. When transmitted to the joint decoder, the lossy versions $U_{1,k}$ and $U_{2,k}$, of $X_k^n$ and $Y_k^n$, , respectively, are obtained. The final reconstructed sequence of information may contain Binary distortion $D_k$. If $D_k \neq 0$, the reconstructed sequence is a lossy version $\hat{X}_k^n$ of $X_k^n$.

\subsection{Channel Model}\label{subchannelmodel}
Consider the transmission of the $k^{th}$ user as an example, when the transmit power $P_{t}$ is fixed, the receive power with the $k^{th}$ user $P_{r,k}$ can be expressed as
\begin{align}
P_{r,k}=\frac{P_{t} h_k}{PL_{i,k}}, \label{eq:num_1xx}
\end{align}
where $h_k$ represents the instantaneous normalized channel gain, $PL_{i,k}$ represents path loss, where $i$ $\in$\{0,1,2\} denoting the S-R link, the S-D link and the R-D link, respectively. The channel is expressed by the A2G channel model, with $i\in\{0,2\}$, and the ground channel model with when $i=1$. According to the literature \cite{AlHourani2014}, the formula for the $PL_{i,k}$ calculation is given as below:
\begin{align}
PL_{i,k}=&\frac{\eta_{LoS}-\eta_{NLoS}}{1+\mathcal{A} \exp(-\mathcal{B}(\theta_k-\mathcal{A}))}\nonumber\\&+20\log_{10}\bigg(\frac{4\pi f_k d_{i,k}}{c}\bigg)+\eta_{NLoS}, 
\label{eq:num_2xx}
\end{align}
where $\mathcal{A}$, $\mathcal{B}$, $\eta_{LoS}$ and $\eta_{NLoS}$  are the environmental parameters. $\theta_k = arcsin\left(\frac{H}{d_{i,k}}\right)$ is the elevation of the UAV, with $H$ being the altitude of the UAV, and $d_{i,k}$ the length of the links. $f_k$ is the carrier frequency, and $c$ is the speed of light.

Then the instantaneous signal-to-noise ratio (SNR) of each link can be calculated by
\begin{align}
\gamma_{i,k}={P_{r,k}}/{N_0}={P_{t}h_k}/({N_0 \cdot PL_{i,k} }), 
\label{eq:num_1xx}
\end{align}
where $N_0$ is the noise power spectral density of the additive Gaussian noise of the receiver. We normalize the average fading channel gain to the unity, and use Nakagami-$m$ fading channel to describe the fading channel variation at the UAV and the users\cite{Khuwaja2018}. It is easy to know that the probability density function of instantaneous SNR is
\begin{align}
p\left(\gamma_{i,k}\right)=\frac{m^m \gamma^{m-1}_{i,k}}{\overline{\gamma}_{i,k}^m\Gamma(m)}\exp(-\frac{m\gamma_{i,k}}{\overline{\gamma}_{i,k}}),
\label{eq:num_1xx}
\end{align}
where the exact value of the factor $m$ can be determined through field measurement campaign in real channels. For simplicity, in this paper, we set it as a parameter. $\overline{\gamma}_{i,k}$ is the average SNR and $\Gamma(\cdot)$ is the gamma function.

\section{Outage Probability Analysis}
\label{sec:analysis}
We derive the relationship between instantaneous SNR and final distortion in two steps, which are, first we identify the achievable rate-distortion region, and then calculate analytically the outage probability.

\subsection{Achievable Rate Region}\label{sana}

We assume a Binary source. A schematic diagram of the problem in the multi-terminal source coding framework corresponding to the system we investigate is shown in \fref{fig:systemmodel2}. 

Let's continue with the $k^{th}$ user. The information sequences $X_k$, $Y_k$, $U_{1,k}$ and $U_{2,k}$ form a Makov Chain $U_{1,k} \rightarrow X_k \rightarrow Y_k \rightarrow U_{2,k}$. For the S-R link, the link rate $R_{0,k}$ should not be less than the information about $X_k$ obtained from the information sequence $Y_k$ in the UAV relay, which is given by the mutual information $I(X_k; Y_k)$. Notice the same rule should apply to the R-D link, the rate $R_{2,k}$ cannot be smaller than the mutual information $I(Y_k; U_{2,k})$.  For the S-D link, $U_{2,k}$ provides helper information, by utilizing the compressed side information $U_{2,k}$ in joint decoding, the rate $R_{1,k}$ only has to be greater than or equal to the mutual information $I(X_k;U_{1,k}|U_{2,k})$.

The requirements for the link rates can be simplified as:
\begin{align}
R_{0,k}  &\geq 1-H_b(\rho_{1,k}) \label{eq:num_5.1} \\ 
R_{1,k}  &\geq H_b(\rho_{1,k} \ast \rho_{2,k} \ast D_k )-H_b(D_k) \\ 
R_{2,k}  &\geq 1-H_b(\rho_{2,k}) 
\label{eq:num_5xx}
\end{align}
where $\rho_{1,k}$ and $\rho_{2,k}$ represent the crossover probabilities with the S-R link and the R-D link, respectively.
 $D_k$ denotes the crossover probability with the S-D link, and $H_b(\cdot)$ denotes the Binary entropy function.
 
\subsection{Outage Probability Derivation}
According to Shannon’s lossy source-channel separation theorem, we can establish the relationship between the instantaneous channel SNR $\gamma_{i,k}$ and its corresponding rate constraint $R_{i,k}$ for $i\in\{0,1,2\}$, as:
\begin{align}
R_{i,k} = {C(\gamma_{i,k})}/{\mathcal{R}_{i,k}}={B\log_2(1+\gamma_{i,k})}/{\mathcal{R}_{i,k}}, 
\label{eq:Rik}
\end{align}
where $C(\gamma_{i,k})$ is the Shannon capacity using the Gaussian codebook, and $\mathcal{R}_{i,k}$ represents the end-to-end rates of joint source channel coding.

Substituting Eq. (\ref{eq:Rik}) into (\ref{eq:num_5xx}), we have
\begin{align}
\rho_{2,k} = H_b^{-1}\bigg(1-\frac{B\log_2(1+\gamma_{2,k})}{\mathcal{R}_{2,k}}\bigg), 
\label{eq:num_6xx}
\end{align}

\begin{figure}[!t]
\centering \includegraphics[width=2.5in]{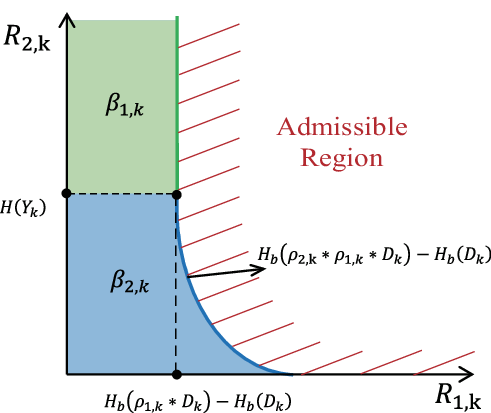}
\caption{The admissible rate region of the $k^{th}$ user; the red solid line indicates the achievable rate region with acceptable distortion $D_k$.}
\label{fig:region}
\vspace{-1em}
\end{figure}

For the achievable rate region of the $k^{th}$ user obtained in Section \ref{sana}, we can further derive the outage probability of the $k^{th}$ user. The outage probability of the $k^{th}$ user can be expressed as
\begin{align}
P_{out,k}=\Pr\{(R_{0,k},R_{1,k},R_{2,k})\in \beta_k\}, 
\label{eq:num_1xx}
\end{align}
where $\beta_k$ represents the admissible rate distortion region of the $k^{th}$ user, as shown in \fref{fig:region}.

The admissible rate distortion region described in Section \ref{sana} needs to be further decomposed for the ease of the calculation. We first focus on the S-D link and the R-D links. For the S-R link, the cross over probability $\rho_{1,k}$ between information $X_k$ and $Y_k$ is regarded as a parameter determined according to  the rate-distortion function, given $R_{0,k}$. Through this method, a two-dimensional achievable rate distortion region can be obtained, as shown in \fref{fig:region}. Moreover, the rate region is divided into two parts corresponding to $\beta_{1,k}$ and $\beta_{2,k}$ to facilitate integral calculation. The outage probability of the $k^{th}$ user can then be expressed as
\begin{align}
P_{out,k}=\Pr\{\rho_{1,k}\in[0,0.5],(R_{1,k},R_{2,k})\in \beta_{1,k} \cup \beta_{2,k}\},
\label{eq:num_1xx}
\end{align}
which can further be decomposed into:
\begin{align}
P_{out,k}=&\Pr\{\rho_{1,k}=0,(R_{1,k},R_{2,k})\in \beta_{1,k}\}\nonumber\\&+\Pr\{\rho_{1,k}=0,(R_{1,k},R_{2,k})\in \beta_{2,k}\}\nonumber\\&+\Pr\{\rho_{1,k}\in(0,0.5],(R_{1,k},R_{2,k})\in \beta_{1,k}\}\nonumber\\&+\Pr\{\rho_{1,k}\in(0,0.5],(R_{1,k},R_{2,k})\in \beta_{2,k}\},
\label{eq:num_8xx}
\end{align}

After several steps of mathematical manipulations, we have:
\begin{align}
P_{out,k}&= 1 + \frac{2\Gamma(m,\frac{m}{\overline{\gamma}_{0,k}})\Gamma(m,\frac{m}{\overline{\gamma}_{2,k}})}{[\Gamma(m)]^2}
\nonumber\\&-\frac{\Gamma(m,\frac{m}{\overline{\gamma}_{0,k}})}{\Gamma(m)}\cdot \int_{0}^{1} p(\gamma_{2,k})
\nonumber\\&\cdot\frac{\Gamma(m,\frac{m\cdot[2^{H_b(\varphi(\gamma_{2,k})\ast D_k)-H_b(D_k)}-1]}{\overline{\gamma}_{1,k}})}{\Gamma(m)}d\gamma_{2,k}
\nonumber\\&-\frac{\Gamma(m,\frac{m}{\overline{\gamma}_{2,k}})}{\Gamma(m)}\cdot \int_{0}^{1} p(\gamma_{0,k})
\nonumber\\&\cdot\frac{\Gamma(m,\frac{m\cdot[2^{H_b(\varphi(\gamma_{0,k})\ast D_k)-H_b(D_k)}-1]}{\overline{\gamma}_{1,k}})}{\Gamma(m)}d\gamma_{0,k}
\nonumber\\&-\int_{0}^{1}d\gamma_{0,k}\int_{0}^{1}p(\gamma_{0,k})p(\gamma_{2,k})
\nonumber\\&\cdot\frac{\Gamma(m,\frac{m\cdot[2^{H_b(\varphi(\gamma_{0,k})\ast  \varphi(\gamma_{2,k})\ast  D_k)-H_b(D_k)}-1]}{\overline{\gamma}_{1,k}})}{\Gamma(m)}d\gamma_{2,k},
\label{eq:num_1xx}
\end{align}
where $\varphi(\gamma_{i,k})$ represents the inverse function of the Binary entropy of $H_b^{-1}(1-\log_2(1+\gamma_{i,k}))$, $\Gamma(\cdot,\cdot)$ stands for the upper incomplete gamma function. It should be noted that the boundaries of the integrals have been properly replaced corresponding to Eq.(\ref{eq:num_6xx}) and no further integral calculation can be performed analytically.
\section{SAC-Based Outage Probability Minimization}\label{sec:sac}
In this section, we design the algorithm based on SAC method. Compared with deterministic strategies, SAC algorithm uses random strategies, which has advantages yielding good performance in continuous actions and state space.

\subsection{Problem Framework}\label{SSAAS}
We model the considered outage probability optimization problem as a Markov Decision Process (MDP). At the time slot $t$, the agent obtains the state $s_t$ by observing the environment and performs an action $a_t$. If the environmental state happens, of which  the occurrence probability is $p(s_{t+1} | s_t, a_t)$, agent will receive a reward $r_{t+1}$ according to reward function. The state, action, and reward functions of the MDP are defined as follows.

$1)$ $States$: The state of the UAV at each time slot $t$ contains the information about the horizontal location of the UAV, which is describe as $s_t = (n_{1,t},n_{2,t})$. 

We consider a constant UAV altitude and only solve the problem of UAV deployment in the two dimensions. Because in the actual scenario, the UAV will usually use the lowest altitude to obtain the best relay performance, optimizing the drone height becomes irrelevant. 

$2)$ $Actions$: The UAV can fly in any direction, and hence its space of action is continuous. Since the action of  each time slot $t$ is denoted by $a(t) = (\Delta n_{1,t},\Delta n_{2,t})$. The action space range is limited, such scenario is equivalent to limiting the flight speed of the UAV.

$3)$ $Rewards$: Since we want to minimize the outage probability of the communication system, we consider incorporating the outage probability into the reward function. At each time epoch t, the reward is calculated based on the
reward function given by
\begin{align}
r_{t+1} 
=\frac{K}{\mu \cdot \sum_{k=1}^K P_{out,k}(t)}, 
\label{eq:num_1xx}
\end{align}
where the constant $\mu$ needs to be tuned empirically.

\subsection{Algorithm Description}\label{RF}
SAC algorithm is an off-policy algorithm developed for maximum entropy RL\cite{Haarnoja2018}. It not only maximizes the long-term return, but also maximizes the entropy of each output action of the strategy, so as to achieve larger exploration, faster convergence, and higher stability. Its optimization objective can be expressed as
\begin{align}
J(\pi) = \mathbb{E}_{(s_t,a_t)\sim \rho^{\pi}}\mathop{\displaystyle\sum}\limits_{t}r(s_t,a_t)+\alpha \mathcal{H}(\pi(\cdot|s_t)), 
\label{eq:num_1xx}
\end{align}
where $\pi$ represents the action function of the agent, the policy. $\rho^{\pi}$ denotes the state-action marginal distribution induced by the policy $\pi$. $\mathcal{H}(\pi(\cdot|s_t))$ is the information entropy of $\pi(\cdot|s_t)$, and $\alpha$ is the temperature coefficient which determines the importance of entropy with respect to the reward, thus controlling the degree of randomness of the policy.

The temperature coefficient needs to be adjusted according to specific factors such as the optimization objective, which imposes some level of difficulties to the training. 
Thus, we use the method of automatically adjusting the temperature coefficient. At this time, the objective function becomes as follow:
\begin{align}
\mathop{\max}\limits_{\pi} &\mathbb{E}_{ \rho^{\pi}}\bigg[\mathop{\displaystyle\sum}\limits_{t}r(s_t,a_t)\bigg]
\nonumber\\\text{ s.t. }  \mathbb{E}_{(s_t,a_t)\sim\rho^{\pi}}&[-\log(\pi_t(a_t|s_t))]\ge\mathcal{H}_{\min}, \forall t ,
\label{eq:num_1xx}
\end{align}
Here, $\mathcal{H}_{\min}$ is entropy target, which is set as $\mathcal{H}_{\min}=-\dim(A)$, the opposite of the dimension of the action space.

\begin{table}[!ht]
\centering
\caption{ System Parameters}
\label{tab:t1}
\begin{tabular}{|c|c|c|c|r|l|} \hline 
 \bfseries Parameters    &  \bfseries Value & \bfseries Parameters    &  \bfseries Value    \\ \hline
$N_0$         & $3.9811\times10^{-14}$ W &$\eta_{LoS}$  &	0.1\\ \hline
$f_1$         & $2$ GHz &$\eta_{NLoS}$ & 21 \\ \hline
$f_2$         & $1.985$ GHz &$P_t$         & $0.5$ W \\
\hline
$\mathcal{A}$           & 4.88 &$m$           & 2 \\ \hline
$\mathcal{B}$           & 0.43 &$\mu$         & $5000$ \\ \hline
\end{tabular}
\vspace{-1em}
\end{table}
\begin{table}[!ht]
\centering
\caption{DRL Hyperparameters}
\label{tab:t2}
\begin{tabular}{|c|c|c|c|} \hline 
 \bfseries Hyperparameters    &  \bfseries Value   & \bfseries Hyperparameters    &  \bfseries Value      \\ \hline
Number of episodes & 300 &
Replay memory size & 10000 \\ \hline
Mini-batch size & 128 &
Learning rate $\lambda_\alpha$ & 0.0003 \\ \hline
Learning rate $\lambda_Q$ &	0.003 &
Learning rate $\lambda_\pi$ & 0.001 \\ \hline
Discount factor $\xi$ & 0.9 &
Activation function & ReLU \\ \hline
\end{tabular}
\end{table}

In the SAC-based algorithm, we use the actor network to generate action policy $\pi$, and the critic network to approximate the soft Q-function to evaluate the performance of the actor network, 
where the soft Q-function is the expected cumulative reward of starting from state $s_t$, taking action $a_t$, and following policy $\pi$.

To avoid duplication of letters, the discount factor in this part is denoted by $\xi$. The soft Q-function parameters $\theta$ and the policy parameter $\phi$ can be trained by minimizing the loss functions (\ref{eq:Q}) and (\ref{eq:pi}) as follow, respectively,
\begin{align}
J_{\mathbf{Q}}(\theta) =& \mathbb{E}_{(s_t,a_t)\sim\mathcal{D}}\bigg[\frac{1}{2}(\mathbf{Q}_\theta(s_t,a_t)-r(s_t,a_t)
\nonumber\\&-\xi\mathbf{Q}_{\bar{\theta}}(s_{t+1},a_{t+1})+\alpha\log(\pi(a_{t+1}|s_{t+1})))^2\bigg],
\label{eq:Q}
\end{align}
\begin{align}
J_\pi(\phi) = \mathbb{E}_{s_t\sim\mathcal{D}}[\mathbb{E}_{a_t\sim\pi}[\alpha\log(\pi(a_t|s_t))-\mathbf{Q}_\theta(s_t,a_t)]],
\label{eq:pi}
\end{align}

The temperature parameter $\alpha$ is adjusted by minimizing the loss function, as follows
\begin{align}
J(\alpha) = \mathbb{E}_{a_t\sim\pi_t}[-\alpha_t\log\pi_t(a_t|s_t)-\alpha\mathcal{H}_{min}].
\label{eq:alpha}
\end{align}

The detailed algorithm is described
in Algorithm~\ref{alg1}.

\begin{algorithm}[!t]
	\renewcommand{\algorithmicrequire}{\textbf{Input:}}
	\renewcommand{\algorithmicensure}{\textbf{Output:}}
	\caption{Optimal Relay Position Identification Algorithm}
	\label{alg1}
	\begin{algorithmic}[1]
		\STATE Initialization the position of UAV;
		\STATE Initialization the experience replay buffer $\mathcal{D}$;
		\FOR{episode $1,2,3...,$}
			\STATE Obtain the initial state s for UAV;
			\STATE Reset episode rewards;
			\FOR{step $1,2,3...,$}
				\STATE choose the action $a_t$ according to the current state $s_t$ the current policy $\pi_t$;
				\STATE  Get the next state $s_{t+1}$ after performing action $a_t$;
				\STATE  Get reward $r_{t+1}$;
				\STATE Store the quadruple $(s_t, a_t, r_{t+1}, s_{t+1})$ to $\mathcal{D}$;

				\IF {Memory counter $>$ Memory Capacity}          
    					\STATE Sample a batch of data in $\mathcal{D}$ randomly;
    					\STATE Update the critic network parameter $\theta$ and the action network parameter $\phi$ by minimizing Equation (\ref{eq:Q}) and (\ref{eq:pi});
    					\STATE Update the  temperature coefficient $\alpha$ by minimizing Equation (\ref{eq:alpha});						\STATE Soft update the parameter $\bar{\theta}$ of the target Critic network;

				\ENDIF
			\ENDFOR
		\ENDFOR
	\end{algorithmic}  
\end{algorithm}

\section{Numerical Results}\label{sec:performance}
We consider a lossy cooperative UAV relaying network consisting of one UAV and two users with their distortion requirements, and the whole scene is distributed in a 20 × 20 km$^2$ area. The flying altitude of the UAV is 500 m, and the initial position is set at being equal to the location of the BS, which is the origin. The locations of the two users are fixed and known. Other parameter Settings in the environment are shown in Table \ref{tab:t1} unless otherwise specified.
All neural networks in the algorithm are implemented based on the Pytorch framework, and the optimizers adopt Adam. In addition, we implement DDPG for comparison. The hyperparameters are shown in Table \ref{tab:t2}.

\begin{figure}[ht]
\vspace{-1em}
\centering \includegraphics[width=3.2in]{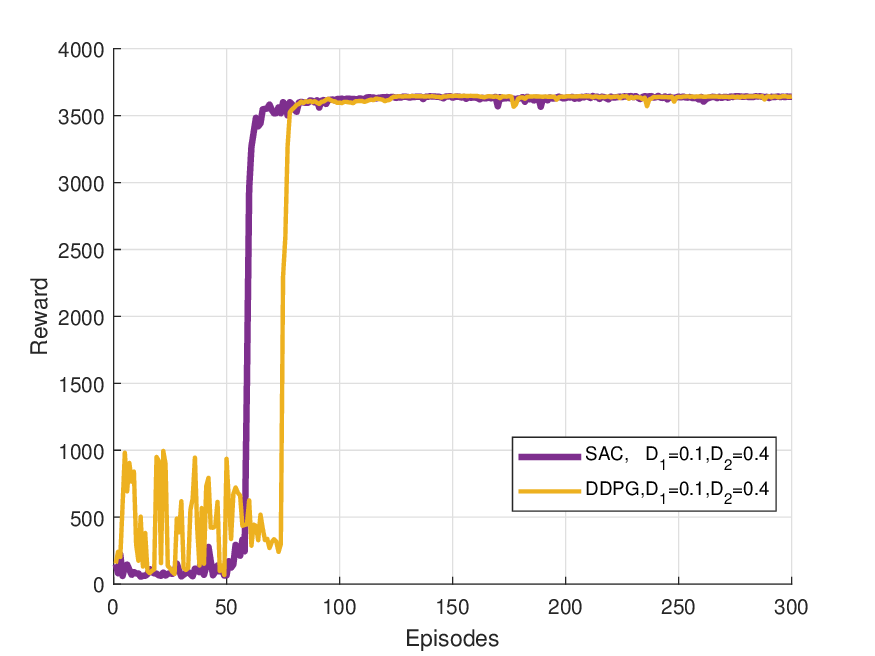}
\caption{The value of reward function over 300 iterations.}
\label{fig:con}
\end{figure} 
\begin{figure}[ht]
\vspace{-1em}
\centering \includegraphics[width=3.2in]{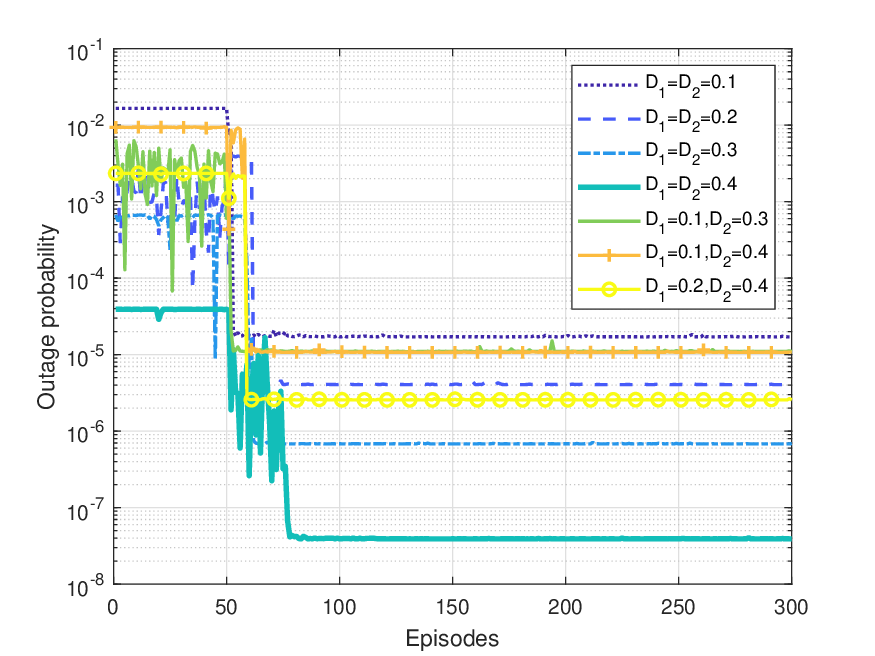}
\caption{The value of reward function over 300 iterations.}
\label{fig:outage}
\end{figure}
\fref{fig:con} depicts the value of the reward function in 300 episodes when $D_1=0.1,D_2=0.3$. The slight instability observed in the SAC based algorithm during the initial phase can be attributed to the algorithm's necessity for extensive exploration to adequately learn the environment and optimize the strategy, which is a typical characteristic of RL algorithms. As the algorithm iterates and learns, the SAC based algorithm gradually converges, exhibiting a high degree of stability. Conversely, the DDPG experiences greater fluctuations and slower convergence.

As shown in \fref{fig:outage}, with the increase of the number of episodes, the outage probability gradually converges into the minimum, which also means that after continuous iterations, the UAV gradually converges to the position yielding the minimum system outage probability. At the same time, it is obvious that the lower the distortion requirement, the larger D value, and hence, the smaller the outage probability of the system. When the distortion requirements are different among users, the outage probability value is mainly limited by the largest distortion requirement.

\begin{figure}[!ht]
\vspace{-1em}
\centering
\subfigure[$D_1=D_2=0.2$.]{\includegraphics[width=2.7in]{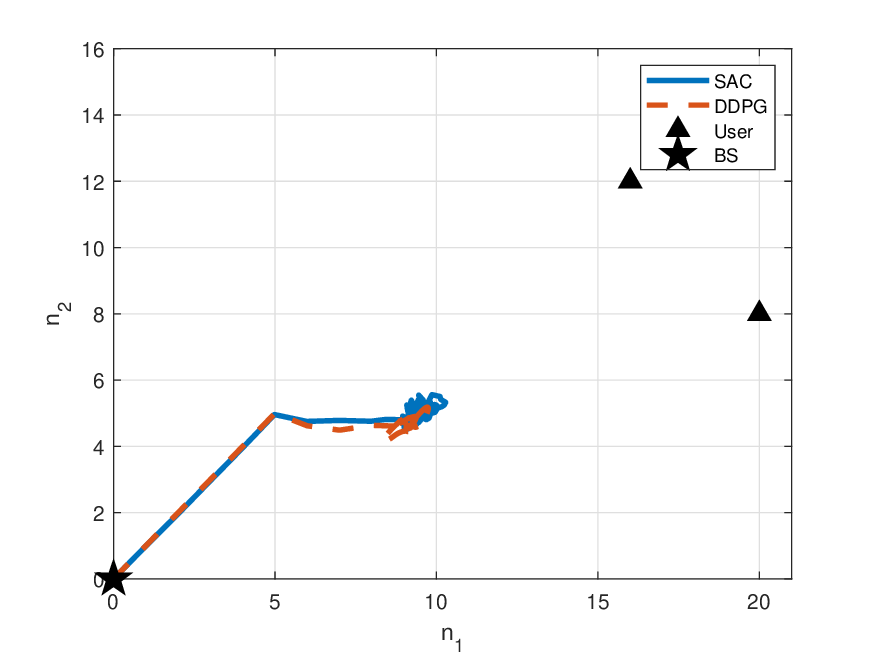}}
\subfigure[$D_1=0.1,D_2=0.3$.]{\includegraphics[width=2.7in]{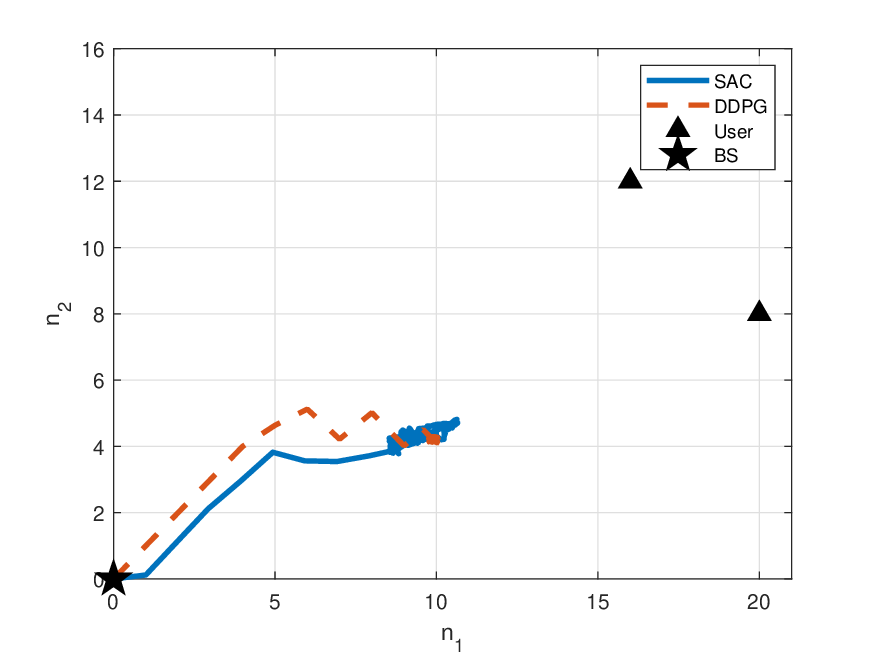}}
\caption{The obtained UAV paths of the two algorithms under different distortion conditions.}
\label{fig:traj}
\end{figure} 

The final trajectory of the UAV is described in \fref{fig:traj}.
It can be seen that compared with DDPG, after the SAC based algorithm converges, the UAV can find a more reasonable trajectory to reach the position with the minimum outage probability.  There will be oscillation in the final trajectory of the UAV. This is because when the UAV flies near the optimal position, the received signal power is already strong, and the influence of horizontal movement on the received power near the optimal position is small, so the UAV oscillates near the optimal position. In future work, the trajectory end point can be stabilized with Kalman filtering, or by a weighted average of historical positions.

\section{Conclusion}\label{sec:conclusion}
In this paper, considering the distortion requirements of the users, we have analyzed the performance of the lossy cooperative UAV relay system, and optimize the UAV trajectory and position with the goal of minimizing the probability of interruption. It has been shown through simulations that the designed optimal relay position identification algorithm can satisfy the requirements of a variety of distortions and it has been shown that the optimal UAV position which minimizes the outage probability can be identified quickly and steadily.

\section*{Acknowledgement}
This paper in part includes intellectual results which the second and the last authors achieved when they were with Japan Advanced Institute of Science and Technology (JAIST) but not published yet, and their extensions. This paper is closely related to the results presented in \cite{Lin2019lossy} which has been published, given the permission by JAIST Research Management Section.  Even though the last author's contributions are on behalf of his current affiliation, they are strictly consistent to the permission for \cite{Lin2019lossy} by JAIST Research Management Section, as well as editorial matters for the readability improvement.

\bibliographystyle{IEEEtran}
\bibliography{myreference}

\end{document}